\documentclass[fleqn,12pt]{article}
\usepackage{latexsym}
\usepackage{graphicx}
\usepackage{amsmath}
\usepackage{amssymb}
\usepackage{amsfonts}
\usepackage{epsfig}
\usepackage{textcomp}	
\usepackage{algorithm, algorithmicx, algpseudocode}
\usepackage{mathptmx}

\usepackage{rotating}

\begin{document}

\title{Variable Neighborhood Search Algorithms for the
multi-depot dial-a-ride problem with heterogeneous vehicles and users}
\author{
 Paolo Detti \thanks{Dipartimento di Ingegneria dell'Informazione e Scienze Matematiche,
University of Siena, Via Roma, 56, 53100 Siena, Italy, e-mail detti@dii.unisi.it}\and   Garazi Zabalo Manrique de Lara   \thanks{Dipartimento di Ingegneria dell'Informazione e Scienze Matematiche,
University of Siena, Via Roma, 56, 53100 Siena, Italy, e-mail garazizml@gmail.com}}

\date{}
\maketitle

\begin{abstract}
In this work, a study on Variable Neighborhood Search  algorithms for  multi-depot dial-a-ride problems is presented. In dial-a-ride problems patients need to be transported from pre-specified pickup locations to pre-specified delivery locations, under different considerations. The addressed problem presents several constraints and features, such as heterogeneous vehicles, distributed in different depots, and heterogeneous patients. The aim is of minimizing the total routing cost, while respecting time-window, ride-time, capacity and route duration constraints.
\newline
The objective of the study is of  determining the best algorithm configuration in terms of initial solution, neighborhood and local search procedures. At this aim, two different procedures for the computation of an initial solution,   six different type of neighborhoods and five local search procedures, where only intra-route changes are made, have been considered and compared. 
 We  have also evaluated an "adjusting procedure" that aims to produce feasible solutions from infeasible solutions with small constraints violations. The different VNS algorithms have been tested on instances from literature as well as on random instances arising from a real-world healthcare application. \\
{\bf keywords}:
Dial-a-ride problem, multi-depot, Heterogeneous users and vehicles, Metaheuristic, Variable Neighborhood  Search.

\end{abstract}

\section{Introduction}
In this work, Variable Neighborhood Search (VNS) algorithms for  the multi-depot dial-a-ride problem (DARP)  with  heterogeneous vehicles and users are presented. In the problem, vehicles have different sizes and features and are located in multiple depots. Users can only use   {\em compatible} vehicles,  and must be transported from a specific origin location to a specific destination location and eventually vice versa. The lengths of the routes assigned to the vehicles cannot exceed given limits.
Quality of service requirements impose the fulfillment of  pickup and delivery time windows and a limited  ride time for each user.

   In this paper,  an extensive computational study is proposed to evaluate different  Variable Neighborhood  Search (VNS) algorithms  for  DARP.  The algorithms are based on the general framework  already introduced in Parragh {\em et al.} \cite{Parragh}, for the single-depot case with homogeneous vehicles and users, and extended  in \cite{DPZomega} to the multi-depot DARP with heterogeneous  vehicles and users in Detti {\em et al.} \cite{DPZomega}.
More precisely, in \cite{DPZomega}, a VNS scheme is  proposed for solving a real-world healthcare application, concerning the  non-emergency transportation of patients in Italy \cite{Agnetis, coppi}. 
In the healthcare application, 
constraints on  arrival and departure times, patient-vehicle compatibility, patients' preferences  and quality of service issues were considered. 

The main  contributions  of this paper are reported below. Starting from the VNS framework developed for the real-world healthcare application addressed in \cite{DPZomega}, new VNS algorithms for the multi-depot DARP with heterogeneous users and vehicles have been designed, presented and tested. The  aim is to evaluate in terms of solution's quality and computational time different $(i)$ algorithms for finding an initial solution, $(ii)$ combinations of  neighborhoods' sets  and   $(iii)$ local search procedures.
Furthermore,  new benchmark multi-depot instances of DARP  with  heterogeneous vehicles and users are presented, generated from the real-world data of the problem addressed in \cite{DPZomega}.
    The  VNS algorithms are evaluated and tested on three sets of benchmark  instances. The first set  has been presented in \cite{Parragh2011} and contains single-depot DARP instances, partitioned into three subsets. The second set has been introduced in  \cite{Braekers} and contains multi-depot instances generated by  the three subsets  presented in \cite{Parragh2011}.   
 As already stated, the third set arises from the real-world healthcare application addressed in \cite{DPZomega}.

    The paper is organized as follows. 
 In Section \ref{sec:lit}, a literature review is presented. 
  In Sections \ref{sec:prob}, a  description of the problem is given. Section \ref{sec:VNS} is devoted to the presentation of the variable neighborhood  search algorithms. In Section  \ref{sec:inst}, the benchmark instances are described. Computational results on the benchmark instances are reported in Section \ref{sec:res}. Conclusions follow in Section \ref{sec:conc}.

\section{Literature review}\label{sec:lit}

DARP is a generalization of the Pickup and Delivery Problem with Time Windows  (PDPTW ) \cite{Bettinelli,Liu, Ropke}. In DARP,  people are transported instead of goods and consequently issues  on the quality of the provided service and timing must be carefully taken into account (through additional constraints or by extra terms in the objective function). For recent surveys on DARP and PDPTW, we refer the reader to Cordeau and Laporte \cite{CorLap}  and Berbeglia {\em et al.} \cite{Berbeglia}.


Promising exact techniques used in the literature for solving DARP, PDPTW and their variants are Branch-and-Price \cite{Bard, Bettinelli, Parragh,Ropke,Savelsbergh} and Branch-and-Cut  \cite{Braekers,Ropke2007}. 
However, due to the difficulty of modeling all the constraints of the real-world problems and of solving large dimension instances, many studies have been also  focused on the development of heuristic and metaheuristic approaches \cite{Ma}.

   Although several studies about DARPs have been proposed in the literature, most of the papers   deal with  a single user type and a homogeneous fleet of vehicles located at a single depot, while few works  address    the with multi-depots, heterogeneous vehicles and users (e.g., see \cite{Braekers},   \cite{Carnes} and \cite{DPZomega}).
   
 In \cite{Braekers},   the multi-depot DARP with heterogeneous vehicles and users is considered, where  quality of service is  limited to  patients' ride times and the route costs  depend on the traveled distance. The authors propose a  deterministic annealing metaheuristic and an exact Branch-and-Cut algorithm based on a the  PDPTW2 formulation of Ropke {\em et al.} \cite{Ropke2007}.
 
  In Carnes {\em et al.} \cite{Carnes}, a multi-depot DARP application coming from an  air-ambulance service context is addressed  (compatibility constraints among patients exist, too, e.g.,  patients cannot be transported with another patient owing to infectiousness). The authors consider a complex cost function for routing costs. The instances handled  in \cite{Carnes} contain no more than 30 requests per day and are solved by a set partitioning formulation. 
 
 As already stated, in  \cite{DPZomega}, a multi-depot DARP is addressed, arising from a real-world healthcare application. For this application, metaheuristic algorithms based on the Tabu Search  (TS) and the Variable Neighborhood  Search (VNS) techniques  are presented and a  Mixed Integer Linear Programming (MILP) formulation is proposed. A performance analysis based on instances generated by real-life data shows that VNS attains the best results in terms of  solution quality and computational times.

Concerning the metaheuristic approaches, the  tabu search and the variable neighborhood search techniques have been widely applied to DARPs. 
 Cordeau and Laporte \cite{CorLap3} first applied TS to a case of DARP with a single depot, where constraints related to vehicles' capacity, route duration and maximum ride time of any user on a vehicle were considered. 
 
Paquette {\em et al.} \cite{Paquette} developed a multicriteria heuristic embedding a tabu search process for solving  a DARP, with a heterogeneous fleet of vehicles and two types of users: ambulatory and wheelchair-bound. In addition to travel costs, the objective function includes  three quality terms, in order to reduce user's inconveniences:  the waiting time  at the pickup node, the waiting time at the delivery node  and user's ride time. The proposed solution procedure combines some features of the tabu search heuristic of Cordeau and Laporte \cite{CorLap3} and of the multicriteria reference point method of Climaco {\em et al.} \cite{Cli}. 
Melachrinoudis {\em et al.} \cite{Melachrinoudisa} proposed a TS heuristic to address a single-depot DARP with soft time windows that arises in a non-profit organization system operating in the Boston metropolitan area. 
To minimize user's inconveniences, they employ an objective function that includes excess riding time, early/late delivery time before service and late pickup time after service.

Beaudry {\em et al.} \cite{Beaudry} implemented a two-phase procedure for the dynamic DARP arising in several large hospitals. Different modes of transportation (e.g. a wheelchair or a stretcher) are considered, and a maximum ride time is given to limit patients' inconveniences.  The first phase consists in generating an initial feasible solution through a simple and fast insertion scheme. In the second phase, this solution is improved with a TS scheme.


Also variable neighborhood search based heuristics have been successfully proposed for solving  DARPs.  Parragh {\em et al.} \cite{Parragh} proposed a VNS heuristic with three neighborhood types for a  single-depot DARP with  homogeneous vehicles, constraints on route duration and time windows and maximum users' ride times. In \cite{Parragh1}, a collaborative scheme has been proposed, integrating the VNS heuristic into a column generation framework, for a variant of DARP,  arising in the Austrian Red Cross context, with driver-related constraints, heterogeneous vehicles and patients. 
Muelas {\em et al.} \cite{Muelas} 
propose a VNS-based algorithm with seven different neighborhood classes (or shakers)
for a  DARP that arises in the area of San Francisco. In their work patients and visitors must be transported from a set of hospitals located around the city to their home addresses (and vice versa), using a fleet of homogeneous vehicles.

Recently, new metaheuristic approaches have been proposed for DARP in  \cite{Braekers} and  \cite{Parragh2013}. As already stated, in  \cite{Braekers}, a deterministic annealing metaheuristic is proposed for the multi-depot  DARP with heterogeneous vehicles and users. In  \cite{Parragh2013}, a hybrid column generation and a large neighborhood search algorithm are presented for solving a single-depot DARP with homogeneous vehicles, maximum user ride times, route duration limits, and vehicle capacity constraints.



\section{Problem Description and Formulation}\label{sec:prob}

In the DARP with multi-depots, heterogeneous vehicles and users, a number of  requests have to be served to transport users from a given origin to a given destination by a fleet of  vehicles, located at geographically distributed depots.

While in the standard DARP, all vehicles have the same capacity and are located at a single depot \cite{CorLap3}, in the heterogeneous multi-depot case, each vehicle, with its own equipment and capacity,  is assigned to a specific depot and must start and end its route at this depot. Users are heterogeneous since they cannot be transported by all the vehicles, but only by the ones with the required equipment.  
The problem consists in assigning the transportation requests to the vehicles and in finding a routing of each vehicle, in such a way  that a total transportation cost is minimized and quality of  service requirements are respected. 

Many constraints can be taken into account, regarding the vehicles'
capacities,  fulfillment  of pickup and delivery time windows, precedence relationships among pickup and delivery locations, patients' preferences and patient-vehicle compatibility, the quality and the timing of the service provided. 
The capacity constraints are related to the number of available seats in the vehicle and to the number
of seats occupied by each users (e.g., in healthcare, the users are patients that may need a stretcher or a wheelchair to be transported, causing an occupation of more than one seat on a given vehicle). The compatibility constraints take into account  the specific setup of the vehicle that can be used to serve a given user typology (for example, a patient on a stretcher can only be  transported by an ambulance). The time windows constraints are related to the users, requiring that each pickup and delivery location need to be reached
within a time interval. 
The precedence relationships state that the patients should be picked up prior to their deliveries.
The quality of service requirements impose that the transportation service fulfills  given indices of the quality of service, such as  limits on the durations of the routes and on the ride times of the users.
In the next section a Mixed Integer Linear Programming (MILP) formulation is given for the problem.

\subsection{A MILP formulation} 
In this section, a three index formulation MILP formulation for the  problem is presented. The MILP formulation extends the formulation proposed by Cordeau \cite{Cor2006} for the standard DARP to the multi-depot case. (The number of the variables of the formulation is reduced by using aggregate variables as proposed in  \cite{Cor2006}.)

First of all, some notation is introduced. Let $G=(V,A)$ be a complete directed graph, where $V=\{{1}, \ldots, {n},{n+1},\ldots, {2n},{2n+1},\ldots,{2n+p}\}$  is the node set and $A=\{(i,j): i,j \in V\}$ is the arc set. 
The  nodes  $P=\{{1},\ldots,{n}\}$ are pickup nodes,  
the  nodes  $D=\{{n+1},\ldots,{2n}\}$ are delivery nodes and the  nodes $DEP=\{{2n+1},\ldots,{2n+p}\}$ represent the depots.  
Hence,  each transportation request $i$ can be denoted by the  nodes $\{i, i+n\}$ in $V$, for $i=1,\ldots,n$.  

The following notation and parameters will be used throughout the paper. 
\begin{itemize}
\item $TR=\{1,\ldots,n\}$   set of transportation requests;
\item $M=\{1,\ldots,m\}$ set of vehicles;
\item $K$ set of vehicle types (defined by the pair vehicle typology and depot in which the vehicle is located);
\item $M_k$, with $|M_k|=m_k$, set of vehicles of type $k$, $k=1,\ldots,|K|$. Hence,  $M=\bigcup_{k=1}^{K} M_k$;
\item $dep(p)$  depot in which   vehicle $p \in M$ is located;
\item $q^{p}$  capacity of  a vehicle  $ p \in M$;
\item $t_{i,j}\ge 0$ travel time for arc $(i, j) \in A$;
\item $d_{i,j}\ge 0$ distance of arc $(i, j) \in A$;
\item $st_i$, {\em service time}, time to load or unload an user at a  node  $i \in P \cup D$;
\item $T_i$ maximum ride time of  the user related to the transportation  request $\{i, i+n\}$;
\item  $K_i$ set of vehicle types that are able (or that are preferred by the user)  to serve the transportation  request $\{i, i+n\}$;
\item $R$ the set of seat types in the vehicles;
\item $C^{p,r}$ number of seats of type $r$ on vehicle $p$;
\item $q_i^{r}$ number of seats of type $r$ that are occupied (if $i$ is a pickup node) or freed (if $i$ is a delivery node) when node $i \in P \cup D$ is visited;
\item $RD$ maximum route duration.
\item $e_i$ beginning of the time window at  node $i$;
\item $l_i$ end of the time window at  node $i$.
\end{itemize}

The time window $[e_i , l_i ]$ of each node $i \in P \cup D$ indicates that a service at node $i$ can only take place between time $e_i$ and $l_i$. A vehicle is allowed to arrive to the location of $i$ before the start of the time window, but it has to wait $e_i$   to begin the load or the unload of the patient. 

 The  MILP formulation employs the following variables.
\begin{itemize}
\item $A_i$ arrival time of a vehicle at  node $i \in P \cup D$;
\item $B_i$ beginning of the service of  a vehicle at  node $i\in P \cup D$;
\item $A^p_i$ arrival time of vehicle $p$ at  depot $i \in DEP$;
\item $B^p_i$ beginning of the service of vehicle $p$ at depot $i \in DEP$;
\item $L_i$ ride time of the user related to node $i$ ;
\item $Q^{p,r}_i$ load of vehicle $p$ leaving node $i$ when user $i$ occupies a seat of type $r$;
\item $x_{ij}^p\in \{0,1\}$ be equal to 1 if vehicle $p$ traversed arc $(i,j)$ and 0 otherwise.
 \end{itemize}

The objective function  reads as follows
\begin{eqnarray}
\min \sum_{i,j \in V}\sum_{p \in M}d_{ij} x_{ij}^p\nonumber
\end{eqnarray}

The constraints of the MILP formulation are reported below. 
\begin{eqnarray}
 \sum_{p\in M} \sum_{j\in V} x_{ij}^p &=&1\; \; \forall i \in P \label{const1}\\
\sum_{j\in V} x_{ij}^p -  \sum_{j\in V} x_{n+i,j}^p&=&0\; \; \forall i \in P, \; p \in M \label{const2}\\
 \sum_{i\in V} x_{ij}^p -  \sum_{i\in V} x_{ji}^p&=&0\; \; \forall j \in V, \; p \in M \label{const3}\\
 \sum_{j\in P} x_{dep(p),j}^p &\le& 1\; \; \forall p \in M \label{const4b}
 \end{eqnarray}
Constraints \eqref{const1} ensure that each request is served exactly once and constraints \eqref{const2} ensure that each origin--destination pair is visited by the same vehicle. Flow conservation is imposed by equalities \eqref{const3}.  Constraints \eqref{const4b}  state that a vehicle $p \in M$ can not leave from the depot $dep(p)$ more than once. 
\begin{eqnarray}
Q^{p,r}_j \ge q_i^{r}  + Q^{p,r}_i - C^{p,r} (1-x_{ij}^p)  \;\; \forall  i,j \in V  \;\; p \in M \;\; r \in R \label{const6}\\
Q^{p,r}_i \le C^{p,r}  \;\; \forall  i \in V  \;\; p \in M \;\; r \in R \label{const6bb}
\end{eqnarray}
 Constraints \eqref{const6}  are load propagation inequalities and Constraints \eqref{const6bb}  limit the load on a vehicle for the different seat types.
 \begin{eqnarray}
A_j \ge B_i + st_i +  t_{ij} - (l_i+ st_i  + t_{ij})(1-x_{ij}^p) \;\; \forall i,j \in P \cup D \;\; p \in M\label{const7} \\
A_j \ge B^p_i + st_i +  t_{ij} - (l_i+ st_i  + t_{ij})(1-x_{ij}^p) \;\; \forall i\in DEP, j \in P  \;\; p \in M\\
A^p_j \ge B_i + st_i +  t_{ij} - (l_i+ st_i  + t_{ij})(1-x_{ij}^p) \;\; \forall i \in D, j \in DEP  \;\; p \in M\\
A_j \le B_i+ st_i + t_{ij} + l_{dep(p)}(1-x_{ij}^p) \;\; \forall i,j \in P \cup D \;\; p \in M\\
A_j \le B^p_i+ st_i + t_{ij} + l_{dep(p)}(1-x_{ij}^p) \;\; \forall i\in DEP, j \in P   \;\; p \in M\\
A^p_j \le B_i+ st_i + t_{ij} + l_{dep(p)}(1-x_{ij}^p) \;\; \forall i \in D, j \in DEP \;\; p \in M\label{const8} 
\end{eqnarray} 
  Constraints \eqref{const7}--\eqref{const8} state that the arrival time at node $j$ must be equal to $B_i + st_i  + t_{ij}$ (to $B^p_i + st_i  + t_{ij}$) if $x_{ij}^p=1$ and $i \notin DEP$ (and $i \in DEP$).
  More precisely,  the constraints  impose that, if $x_{ij}^p=1$,  $A_j$ 
  can not be bigger than $ B_i + st_i  + t_{ij}$ or $ B^p_i + st_i  + t_{ij}$, depending if $i \notin DEP$ or not, respectively. In the constraints, $l_{dep(p)}$ is the end of the  time window at the depot of vehicle $p$.
 
\begin{eqnarray}
L_i&=& B_{n+i}  +st_{n+i}  - (B_i +st_i )\;\; \forall  i \in P  \label{const10}\\
t_{i, n+i}+st_{n+i} \le L_i&\le& T_i \;\;  \forall i \in P  \label{const11}\\
e_i\le  B_i &\le& l_i \;\; \forall  i \in P \cup D \label{const12}\\
A^p_{dep(p)} - B^p_{dep(p)} &\le& RD \;\;\forall  p \in M\label{const15}\\
A^p_{dep(p)} - B^p_{dep(p)} &\ge& 0 \;\; \forall  p \in M \label{const14bis}
\end{eqnarray}
Equalities \eqref{const10} define the ride time of each user. Time window and maximum ride time compliance are ensured by \eqref{const11} and \eqref{const12}, and Constraints \eqref{const15} limit the maximum length of each route. 

  Constraints \eqref{const14bis} state that the arrival of a vehicle to the depot must be after the beginning of the service, corresponding to the departure time of the vehicle.  
Finally, the compatibility constraints between vehicles and patients and  the patient preferences (constraints \eqref{const19}), and  the variable domains can be written as
   \begin{eqnarray}
x_{i,j}^p =0 \;\; i \in P \cup D \; \;  \forall p \in M_k :  \; k \in K \setminus K_i \label{const19}\\
A^p_i, B^p_i, L_i,  Q^p_i \ge 0\; \; \forall i \in V  \;\; p \in M\\
x_{ij}^p \in \{0,1\}\;\; \forall i,j \in V  \;\; p \in M.
\end{eqnarray}

\section{Variable Neighborhood Search algorithms}\label{sec:VNS}
In this section, the general algorithmic  framework used to develop the VNS algorithms is presented.   The framework  is similar to those  presented in \cite{Parragh}   and  \cite{DPZomega}. However,  an algorithm for the initial solution, new neighborhoods  and local search procedures are proposed, with the aim of finding the most promising configurations in terms of solution quality and computational time. The algorithms will be presented into detail and tested in Section \ref{sec:res}.

The general  algorithmic framework has two main steps. In the  step 1, first  an initial solution, $s_0$, is generated. Then, $s_0$ undertakes a local search step, yielding $s$, i.e., the first incumbent solution. The second step of the algorithm is an iterative step. At each iteration $h$,    a new  solution, $s'$, is created in the  neighborhood $N_h(s)$ by a shaking procedure. Afterwards a local search step is applied to $s'$ yielding $s''$. If $s''$ is {\em quasi}-feasible, i.e., only violates one  constraint type of a small amount, a new local search phase (called {\em Adjusting} procedure) is applied to $s''$, in order to get a feasible solution. 
A scheme of the general framework is reported in Algorithm \ref{algo:vns}.

\begin{algorithm}
\caption{Scheme of the Variable Neighborhood Search algorithm}\label{algo:vns}
{\scriptsize
\begin{tabbing}
{\emph{  \textbackslash \textbackslash \; Initial Solution  }} \\ 
Generate the initial solution $s_{0}$;\\
Apply local search to $s_{0} $  yielding $s$;\\
{\bf{If}} $s$ is feasible. Set $s_{best}:=s'_{best}:=s$. {\bf{Else}} set $s_{best}:=s'_{best}:=+\infty$;\\
Set $h:=1$;\\
{\bf {Repeat}}\\
$\{$
{\emph{  \textbackslash \textbackslash \; Shaking  }} \\ Randomly choose $s'$ in $N_h(s)$;\\
\quad \bf{If} $ f_1(s)  < 1.02 f_1(s') $ \\
\quad \quad {\emph{  \textbackslash \textbackslash \; Local Search  }} \\
\quad \quad Apply local search procedure(s) to the routes in $s'$ that have been changed yielding $s''$;\\
\quad {\bf{else}}
\quad  Set  $s'':=s' $; \\
 \quad {\bf{If }} $s'' $  is feasible and $f(s'')<f(s_{best})$\\
  \quad\quad Set $s_{best}:=s'' $ and  $s:=s'' $ and $h:=0;$\\
\quad {\bf {Else}}\\
\quad\quad{\bf{If }}$f(s'')<f(s)$\\
\quad \quad \quad  Set $s:=s''$ and set $h:=0$;\\
\quad\quad\quad  {\emph{  \textbackslash \textbackslash \; Adjusting procedure  }} \\ 

\quad\quad\quad {\bf{If}} $s''$ only violates capacity, (or ride time or time windows) constraint and $q(s)<q_{max}$ $\left( or t(s)<t_{max} or w(s)<w_{max} \right)$\\
\quad\quad\quad \quad
 Apply the adjusting procedure to $s''$ yielding $s_{adj}$;\\
\quad\quad\quad\quad {\bf{If}} $s_{adj}$ is feasible and $f(s_{adj})<f(s'_{best})$\\
\quad\quad\quad\quad\quad Set $s'_{best}:=s_{adj}$; \\
\quad Set $h:= (h \bmod h_{max} )+1$;\\
$\}$ 
{\bf{Until}}  $It_{max}$ iterations are reached \\
{\bf{If}} $f(s_{best})<f(s'_{best})$ return $s_{best}$ {\bf{Else}} return $s'_{best}$.

\end{tabbing}
}
\end{algorithm}

\subsection{Evaluation Function}
As in  
\cite{DPZomega}, \cite{CorLap3} and  \cite{Parragh}, the evaluation function  $f(s)$ used by the VNS algorithms  has two main terms and is equal to $f(s) = f_1(s) + f_2(s) $, where $s$ is a problem solution. 
Function $f_1(s)$  contains the objective function of the problem, i.e., the total distance travelled by all the vehicles. 
The term $f_2(s)$ is a penalization component of the form:
$f_2(s) = \alpha t(s) + \beta w(s) + \gamma q(s)+\tau d(s) $,
where $t(s)$, $ w(s)$, $ q(s)$ and $d(s)$ represent the total violation of solution $s$ with respect to the constraints on patient ride times, time windows, vehicle capacities and route duration, respectively, and $\alpha$, $\beta$,  $\gamma$ and $\tau$ are  positive penalty coefficients. 
Let $p(i)$ be the vehicle serving node $i$  in the solution $s$.  
In more detail, violations are calculated as follows:
\begin{itemize}
\item $t(s) = \sum_{ i\in P}  (L_i - T_i)^+$;
\item $w(s)=\sum_{i \in P \cup D} (A_i- l_i)^+ + (e_i - A_i)^+$. Recall that early arrival is allowed, but the vehicle has to wait until the start of the time window to begin the load or the unload service. 
\item $ q(s)= \sum_{r \in  R}\sum_{ i \in P \cup D}  (Q^{p(i),r}_{i}-C^{p(i),r})^+$. 
\item $d(s) =\sum_{p\in K} ((A^p_{dep(p)}-B^p_{dep(p)})-RD)^+.$
\end{itemize}
Initially, coefficients $\alpha$, $\beta $, $\gamma$ and $\tau $ are set to given values  $\alpha_0$, $\beta_0 $, $\gamma_0$ and $\tau_0 $, respectively, and each time a new incumbent solution is found are modified by a factor of $1 + \delta$. In the VNS, $\delta $ is randomly chosen between $0.05$ and $0.1$, and  changes every time a new incumbent solution is found.

\subsection{Initial Solution}\label{sec:initial}

Two heuristics have been used to generate the initial solution $s_0$ in the algorithms. 

The first heuristic (called Heuristic 1) is a constructive heuristic already used in \cite{DPZomega} and  works as follows. The transportation requests  are first sorted according to the pickup earliest time, $e_i$. According to this ordering, the requests  are assigned and inserted in the routes: first  the pickup node is inserted  and right after the delivery node. A request 
can be assigned to a route only if the compatibility constraints  are respected. Capacity or ride time violations are not considered during this insertion phase. 
The assignment of the request to the routes is performed  according to a minimum distance criterium. More precisely, a request is assigned to the route minimizing the average between the distance from the destination of the last request on the route (if the route is still empty the depot destination is taken instead) to the origin of the  request to assign, and the distance from the destination of the  request to assign to the depot. 
If an assignment does not  satisfy  the time window constraints or pre-specified limits on the length of the route, the insertion is not performed and  the next route that minimizes the  distance criterium, as described above, is considered. If, at the end,  the request is not assigned to any route, the request is assigned to the route  
that produces the smallest increase on the route length (without regarding the time window constraints).

The second heuristic (called Heuristic 2), computationally more expensive, works as follows. The transportation requests are first randomly sorted. According to this ordering, each request is assigned and inserted in all the routes in the best possible position, i.e., the position that minimizes the objective function.  This procedure is repeated $1000$ times, and the solution with the smaller value of the objective function is chosen as initial solution.

\subsection{Shaking}
Six different types of neighborhoods have been considered, called  {\em  swap}, {\em  move},  {\em  repairing move}, {\em  chain}, {\em eliminate} and {\em intra-move}. Neighborhoods swap, chain, and  move  have been introduced  in  \cite{Parragh} and  \cite{Parragh2011}, respectively. Repairing move have been introduced  in \cite{DPZomega}. Finally, the neighborhood eliminate is an adapted version of that used in  \cite{DPZomega},  and intra-move is a new neighborhood developed specifically for the problem considered in this paper. 

The VNS algorithms preseted into detail in Section  \ref{sec:res} contain different subsets of the above neighborhoods. In what follows, a description of each neighborhood is given.

\subsubsection{Swap }
In the neighborhood swap, two sequences of requests belonging to two different routes are exchanged. The two sequences have a length not bigger than $h$, where $h$ is  the size of the current neighborhood. 
Hence, a swap of size $h=2$ consists in exchanging two sequences not longer than two. 
More precisely, the two sequences are exchanged, as follows. First  two routes are  randomly selected. Then, 
the lengths of the sequences in the two routes  to be swapped are randomly chosen. (These lengths must be at most  the size of the neighborhood  $h$). Then, the first nodes of the two sequences  are randomly chosen. Due to the compatibility and preference constraints, all requests of the first (second) sequence must be compatible with the vehicle  of the second (first) route. Otherwise, the whole process will start again. After the selection of the routes and of the sequences, the requests of each sequence are removed from their original route and inserted in  the best possible position in the destination route (respecting the order of the sequence). The allocation of the requests of each sequence is done one by one,  first inserting the pickup and then inserting the delivery. 
The insertion is performed in the location that minimizes the evaluation function. 

\subsubsection{Move } In    the neighborhood move, a number of transportation requests are randomly selected and inserted in  randomly chosen routes, respecting compatibility and preference constraints. The number of transportation requests to move is randomly chosen between $1$ and the size of the current neighborhood, $h$. 
The insertion of the requests is performed as  in swap. Hence, a move of size $h=3$ consists in inserting at most three transportation requests each of them on a different randomly selected route. 

\subsubsection{Repairing  move }
Repairing  move  basically works as move, but it applies  to infeasible solutions only. In   repairing move, the transportation requests to move are randomly selected only from infeasible routes and inserted, as in move, in  randomly chosen routes. The maximum number of transportation requests to move is randomly chosen between $1$ and the size of the current neighborhood, $h$. 
The insertion of the requests in the randomly selected routes is performed as  in move and swap.

\subsubsection{Chain }
In  chain, a sequence of requests is first randomly selected as in  swap from an {\em initial}  route, say $r_0$. Then,  the sequence is inserted, as in swap, in a randomly selected  {\em destination}  route,  compatible with the requests of the initial sequence. The above procedure is repeated on the destination route (that becomes the new initial route). The procedure is repeated at most $h$ times, where $h$ is the neighborhood size. Hence, a chain of size $h=3$ consists in moving a random sequence of  length at most three from its initial route to a destination route. Then, from this destination route, another sequence of  length at most three is moved to another destination route. Finally, from this third route again a sequence of  length at most three is moved to a fourth route.

\subsubsection{Eliminate}
Eliminate generates new solutions by eliminating one existing route. The route with the biggest ratio between the total cost of the route and the number of requests in the route is eliminated. All the requests on the route are eliminated from the route an inserted on randomly chosen compatible routes. The insertion of the requests in the randomly selected routes is performed as in move and swap. 
 \subsubsection{IntraMove}
 IntraMove generates new solutions by removing some requests and re/inserting them in the same routes. A number of requests is randomly chosen. These requests are eliminated and re-inserted in the same routes. The insertion is made as in swap.
 
\subsubsection{Neighborhood Order}

In the VNS algorithm, first  the simplest neighborhoods are applied and afterwards  the more complex and time consuming ones. The IntraMove neighborhood is done at the end of the sequence. More precisely, the neighborhoods are applied with the following order: $S1-RM1-M1-C1-E1-S2-RM2-M2-C2-E2-\ldots-S_{h_{max}}-RM_{h_{max}}-M_{h_{max}}-C_{h_{max}}-E_{h_{max}}-IM$. Where $h_{max}$ is the {\em maximum size} of the neighborhood (in the experiments $h_{max} \in \{2; 3; 4; 6\}$).
Let $SN$ be the subset of neighborhoods employed in a the shaking phase of a given VNS algorithm. During its execution, the neighborhoods in $SN$ are applied according to the above order.

Furtheremore, in all the algorithms, in order to avoid unnecessary operations,  only the routes that have been changed by the shaking procedure are involved in the local search.

\subsection{8-step evaluation scheme}\label{sec:8-step}

In order to set the beginning of a request in a route in the best possible way and to minimize the total duration of the route, the {\em 8-step evaluation scheme} introduced    in \cite{CorLap3} and used in \cite{Parragh}, too, is considered and applied.  Given a route, it employs the idea of forward time slack, $F_i$, where $F_i$ is defined as  the largest increase in the beginning of the request at node $i$ that does not cause any violation.

\subsection{Local Search}
Five different   local search procedures have been developed and tested. The first three procedures (called Local Search 1, Local Search 2 and Local Search 3) can be used in alternative, the fourth and the fifth (called Focus Local Search and Focus Local Search with 8-step)  can be  used in combination with one of the first three procedures. The local search procedures are described in the following.

\subsubsection{ Local Search 1}\label{sec:ls1}
Local Search 1 is a procedure in which only intra-route changes are made. The local search procedure is very similar to the one proposed  in \cite{Parragh}, but it performs a smaller number of changes and is computationally less expensive, as explained in the following. 
Given a route changed by the shaking procedure, 
the nodes corresponding to each transportation  request  are iteratively removed and inserted in new positions on the route, if the evaluation function is improved. 
More precisely, given a route,
the first pickup node and its delivery node are removed from the route. 
Then, the pickup node, if the request is an inbound request, or the delivery node, if it is an outbound request are inserted in the best possible position according to the time window constraints.  (Observe that, while in the procedure proposed in \cite{Parragh} such a  node can be further moved during the search process, in our procedure the position of this node is not changed anymore.)
Afterwards, the remaining node is inserted 
right after (if it is a pickup node) or right before (if it is a delivery node) the first assigned node. If this change improves the evaluation function, then the nodes are fixed in the new positions, otherwise
the procedure attempts to move the remaining node to  the next possible position, increasing the distance from the first assigned node. 
This is repeated until the evaluation function is improved or no other possible position exist for the remaining node. If no improvement is found the request is reinserted in the original position. The procedure continues considering the nodes of the next request on the route. The local search ends when the procedure described above is applied to all the routes changed by the shaking procedure. 

\subsubsection{ Local Search 2}\label{sec:ls2}
As in Local Search 1, in this procedure given a route, the first pickup node and its delivery node are removed from the route. Then, the pickup node, if it is an inbound request, or the delivery node, if it is and outbound request, is inserted in the best possible position according to the time window constraints. 
Afterwards, the remaining node is inserted in all the possible positions  right after (if it is a pickup node) or  before (if it is a delivery node) the first assigned node. The remaining node is assigned to the position that minimizes the evaluation function. 
Then, the procedure continues with the next request on the route. The procedure ends when all the routes in the solution have undertake this local search step. 

Observe that, Local Search 2 can be computationally more expensive than  Local Search 1, since it attempts to assign  the remaining node of each request in all possible positions.

\subsubsection{Local Search 3}\label{sec:ls3}
Local Search 3 is very similar to   Local Search 2, but it performs a smaller number of moves. In fact, as in  Local Search 2, first the critical node is inserted in the best possible position according to the time window constraints. Afterwards, it attempts to assign the remaining node in all the possible positions  which do not violate the time window constraints. 
All the insertions are made as in Local Search 1, first inserting the critical node and then the remaining node. 

\subsubsection{Focus Local Search}\label{sec:fls}

Focus Local Search operates on the sequences of transportations requests, belonging to the same route, starting and ending with an empty load. In the procedure,  
first the routes containing such a sequences  are identified. Then, the sequences are removed from the corresponding routes. The requests involved in the sequences are re-inserted in the same routes following the same insertion procedure used in swap or chain. 

\subsubsection{Focus Local Search with 8-step }\label{sec:fls8}
This local search procedure basically operates as the  Focus Local Search, but just after an insertion of a request the 8-step evaluation scheme (see Section \ref{sec:8-step}) is applied to the (partial) current solution in order to minimize the total duration of the route. Hence, this procedure is computationally more expensive than the Focus Local Search. However, the overall computational increase is limited, since only  sequences of transportations requests  starting and ending with an empty load are involved in the process.


\subsection{Adjusting Procedure}
This procedure tries to produce feasible solutions by perturbing quasi-feasible solutions. It has been first introduced in \cite{DPZomega}, and is effective in instances in which the ratio $n/m$ is small (i.e., when the number of vehicle is large with respect the number of transportation requests).
 
The procedure involves  incumbent solutions $s$, found during the shaking procedure  only violating a single constraint's type, i.e., capacity, ride time or time window, and in which the violation is below a given threshold. 
The adjustment procedure chooses, one by one,  requests violating the constraint and assigns them either $(i)$  at the beginning or at the end of other already used compatible  routes, or $(ii)$ to  empty compatible  routes, if any. 
The assignment is performed to the route minimizing the evaluation function.
The possibly feasible solution obtained by the adjusting procedure   is then compared with the best feasible solution obtained so far by the algorithm.

  \section{Benchmark instances}\label{sec:inst}
  The experimental campaign has been performed on three sets of benchmark  DARP instances.
      
  The first set  has been proposed in \cite{Parragh2011} and contains three subsets of single-depot DARP instances, denoted as $U$, $E$ and  $I$ respectively. 
  These instances, denoted in  \cite{Parragh2011} as set $I$, have been generated from the $a$ instances provided by Cordeau \cite{Cor2006}, in which heterogeneous real-life aspects  arising at the Austrian Red Cross have been introduced. 
  The set  $U$ contains  instances with  a single user/resource type and a homogeneous vehicle fleet, the set  $E$ includes four resource types with a homogeneous fleet of vehicles, and the  set $I$ contains four user/resource types and a heterogeneous fleet of vehicles.
More precisely, the instances in $I$  contain up two types of vehicles and   four types of resources: staff seats, patients seats, stretchers and wheelchair places. The users are characterized by the   type of seat they require on a vehicle and are of four types: accompanying person, seated patient, patient on on a stretcher or in a wheelchair.
Accompanying persons may use staff seats, patients seats or may sit on the stretcher. Seated patients may use a patient seat or may sit on a stretcher. Finally, users transported on a stretcher or in a wheelchair may only use the corresponding places.  Table \ref{tab:seat} summarizes  the pair patient/seat type that are compatible. 
\begin{table}
\centering

\begin{tabular}{ | l | l | l | l | l |}
\hline
Passenger type & Staff seat & Patient seat & Stretcher & Wheelchair \\ \hline
Staff & x & x & x &  \\ \hline
Seated & &x & x &  \\ \hline
Stretcher  & &  & x &  \\ \hline
Wheelchair &  &  &  & x \\ \hline

\end{tabular}
\caption{Table of compatibility patient type vs seat type}\label{tab:seat}
\end{table}

The second set contains 24 multi-depot instances introduced in  \cite{Braekers}. The instances have  been generated  introducing four vehicle depots in the instances of the set $I$ (presented in \cite{Parragh2011}). The four  depots are respectively located  at the  coordinates $[-5; -5]$, $[5; 5]$, $[-5; 5]$ and $[5; -5]$. The vehicles are assigned to the depots in a round robin way.

  The third set contains   instances introduced in  \cite{DPZomega}, extracted from real-world data.
  The instances are partitioned into two subsets, called {\em large} and {\em small} instances, and include   transportation requests operated in a day by a Local  Health Care Agency in  Tuscany, an Italian region. In these instances, the users are of three types:  seated patient, patient on  a stretcher or in a wheelchair. The vehicles belong to four different types, ambulance, bus, car and equipped vehicle, distributed over up to 17 depots. Ambulances and  equipped vehicles can be transported at most one patient at time, while buses and cars have a maximum capacity of 8 and 3 seats, respectively.  
  In each instance, patients  can be transported only by a subset of vehicles (identified both by the patient's conditions and by  the patient's preferences).
 In general,  patients  on  a stretcher can be only transported by an ambulance,  patients on a wheelchair can be transported by equipped vehicles or  ambulances, while all other patients can use all the vehicles. 
  The large instance set contains 14  instances with the following characteristics: 
 Instances 1--4 have 80 transportation requests and 60 vehicles, instances 5--9 have 100 transportation requests and 75 vehicles, and instances 10--14 have 100 transportation requests and 80 vehicles. 
The small instance set contains  10 instances, in which the number of requests and  vehicles are in the range 10--35 and 4--20, respectively.   
 The service time to load or unload a patient  is 15 minutes in the large instances and 10 minutes in the small instances. In all the instances, the maximum ride time for a transportation request is  given by the traveling time from the pickup and delivery location multiplied by 2. Although the instances have been tested in \cite{DPZomega}, too,  by using    a complex objective function composed of several terms (including the length  and the duration of the routes, waiting times of the vehicles, etc.), here we only consider the total traveled distance as objective function, as for the first two sets of benchmark instances introduced above. In all the distances a route duration limit  has been set, too. The instances of the third set are available at {\em http://www.dii.unisi.it/\texttildelow detti/darp-instances}.
 


\section{Computational results}\label{sec:res}
Two test phases have been performed. In the first phase, a preliminary computational campaign has been performed, in which twelve VNS algorithms have been developed and tested on the instances of the second set. In the second phase, new VNS algorithms have been generated and tested on all the multi-depot instances (Section \ref{res:multidep}) and on the single-depot instances (Section \ref{sec:resparragh}), starting  from the best algorithms detected in the first phase.

\subsection{Preliminary computational results}

In order to asses the best neighborhood sets,  local search and improvement procedures  a preliminary test phase has been  performed, in which many VNS algorithms have been  tested. At this aim,  the  24 multi-depot benchmark instances introduced in  \cite{Braekers} have been used (i.e., the instances of the second set).

Twelve VNS algorithms have been developed, characterized by different  neighborhood sets,  maximum size values  and     local search procedures.
The features of the twelve VNS algorithms are summarized in Table \ref{tab:PrelDesc}. Hence, as an example, referring to the general algorithmic scheme of Algorithm \ref{algo:vns}, Algo. \#  1 employs $(i)$ a shaking phase containing  all the neighborhoods except IntraMove,  $(ii)$ a local search phase with Local Search 1 and the Focus Local Search, and $(iii)$ a maximum size of the neighborhood ($h_{max}$) equal to 6. In all the twelve algorithms, the initial solution is found by applying the Heuristic 1 (see Section \ref{sec:initial}), and $\alpha_0=100$, $\beta_0=1$, $\gamma_0=10000$, $\tau_0=1$, and $IT_{max}=15000$ have been set.


\begin{table}
\centering
\begin{tabular}{|l|l|l|l|l|l|l|l|l|l|l|l|l|}
\hline
Algo.  \#     & 1 & 2 & 3 & 4 & 5 & 6 & 7 & 8 & 9 & 10 & 11 & 12 \\ \hline
Swap       & X & X & X & X & X & X & X & X & X & X  & X  & X  \\ \hline
Rep Move   & X & X & X & X & X & X & X & X & X & X  & X  & X  \\ \hline
Move       & X & X & X & X & X & X & X & X & X & X  & X  & X  \\ \hline
Chain      & X & X & X & X & X &   & X & X & X & X  &    &    \\ \hline
Eliminate & X & X & X & X &   & X &   &   &   &    &    &    \\ \hline
IntraMove  &   &   & X & X & X &   &   &   &   &    &    &    \\ \hline
LS1     & X & X & X & X &   &   &   &   &   &    &    &    \\ \hline
LS2     &   &   &   &   & X & X & X & X & X & X  & X  & X  \\ \hline
Focus LS    &   & X &   & X & X & X &   & X &   & X  &    & X  \\ \hline 
MaxSize    & 6 & 6 & 6 & 6 & 3 & 3 & 2 & 2 & 4 & 4  & 4  & 4 \\ \hline
\end{tabular}
\caption{Preliminary results}\label{tab:PrelDesc}
\end{table}

Table \ref{tab:PrelAll}, for each algorithm, reports the average results on the 24 instances. Ten runs of the algorithms on each instances have been performed, resulting in 240 runs for each algorithm.  For each algorithm and instance, the results of the best 5 runs out of 10 (in terms of objective function value) have been only considered for  Table \ref{tab:PrelAll}. In the second column, ``av\_cost" is the average objective function value obtained by each algorithm on  the best 120 runs (out of 240). For each algorithm, in column 3, ``min\_cost"  reports  the average  best objective solution value (on the 24 instances)   found on the 5  best runs,  while,  in column 4, ``\# opt" contains the number of instances, out of 24,  in which  an optimal solution is found (given in \cite{Braekers}). Column 6 reports the average computational time on the 120 best runs of the algorithms. Finally, columns 7 and 8 contain the percentage average and minimum gap, denoted as  ``av\_gap'' and ``min\_gap'', respectively. More precisely, ``av\_gap'' is the average, on the 24 instances,  of the gaps $av\_gap_i$, one for each instance $i$, between the average objective function value, $av\_cost_i$ (on the 5 best runs) and the  optimal solution value $opt_i$,  given   in  \cite{Braekers}. $av\_gap_i$ is computed as $(av\_cost_i-opt_i)/opt_i\times 100$. 
A similar definition holds for ``min\_gap'', where, for each instance $i$, the best  objective function value  on the 5 best runs is used instead of  $av\_cost_i$.



We note that, the best results in terms of solution quality (including the number of times in which an optimal solution is found)  are attained by Algorithms 9 and 12. 
In fact, the ``min\_gap'' values for are Algorithms 9 and 12 are 3.73\% and 3.88\% while the other algorithms have ``min\_gap'' values always bigger than 4\%. Moreover, Algorithm 9 is one of the fastest,  with a computational time of  150 seconds, on average. Observe that  algorithms 1-4 have sensibly higher computational times, mainly due to the neighborhood Eliminate, computationally expensive. 

In terms of optimality, Algorithm 2 finds an optimal solution on 4 of the 24 instances (while the others algorithms  no more than 3), however  it attains on average higher gap values than Algorithms 9 and 12.

\begin{table}
\begin{center}
\begin{tabular}{|c|c|c|c|c|c|c|}
\hline
Algo. & av\_cost & min\_cost    &\# opt & time& av\_gap& min\_gap\\
\hline
1&666.47&649.64& 3 &277.70&6.60&4.08\\
2&667.01&649.45& 4 &308.79&6.63&4.21\\
3&665.44&651.55& 3&276.17&6.50&4.45\\
4&664.87&651.87& 3 &322.12&6.35&4.46\\
5&666.18&652.43& 2 &159.91&6.80&4.60\\
6&674.41&655.78& 2 &309.43&7.78&5.03\\
7&681.11&662.97& 1 &117.60&9.09&6.39\\
8&679.55&661.29& 1 &128.60&8.83&6.15\\
9&655.17&646.65& 3 &149.78&4.93&3.73\\
10&663.83&652.37& 2 &176.79&6.23&4.51\\
11&666.88&649.04& 1 &154.77&6.65&4.13\\
12&658.61&647.6& 3 &172.26&5.44&3.88\\
\hline
 \end{tabular}
\end{center}
\caption{ Preliminary results.}\label{tab:PrelAll}
\end{table}

\subsection{Results of the second test phase on multiple depot instances}\label{res:multidep}
Starting from the best algorithms detected in the previous section a new computational campaign has been performed on  multiple depot instances (i.e., the instances of the second and third  sets, introduced in Section \ref{sec:inst}). 
More precisely, new 6 VNS algorithms have been developed and tested in this phase. 
The features of the algorithms are summarized in Table \ref{tab:algo2}.
Hence, Algorithms 13 and 14  have the same neighborhood set of algorithms 9 and 12, respectively, but the local search procedure LS3 in used in place of LS2.
Algorithms 15 and 16 have the same neighborhood set of algorithms 9 and 12, respectively, too, but they only use the Focus Local Search (see Section \ref{sec:fls}). 
Finally, Algorithms 17 and 18 have the same neighborhood set of algorithms 15 and 16,  but they only use the Focus Local Search with the 8-step evaluation scheme (see Section \ref{sec:fls8}). 
  
   In Tables \ref{tab4} and \ref{tab5}, the results of the 6 algorithms are reported. The solution provided by Heuristic 1 has been used as starting solution. 
 For each instance  ``av\_gap'' (``min\_gap') is the gap between the average (minimum) objective function value, ``av\_cost''   (``min\_cost"),      on the 5 best runs and the optimal solution ``opt", given in  \cite{Braekers}.  ``av\_gap'' (``min\_gap') is computed as $(av\_cost-opt)/opt\times 100$ ($(min\_cost-opt)/opt\times 100$).

  Observe that, in terms of solution quality, Algorithms 13, 15 and 17 attain the best results, while Algorithms 16 and 18 require the smallest computational times. In fact, the complex  neighborhood Chain is not used in Algorithms 16 and 18. In terms of solution quality, Algorithms 13 and 17 attain the best results with average (minimum) gaps of  4.81\% (3.59\%) and 4.8\% (3.65\%), respectively.

  In order to evaluate how  the initial solution influences the VNS technique, Algorithms 13, 15 and 17 has been executed starting from a  solution provided by Heuristic 2 (introduced in Section \ref{sec:initial}).
  The results, reported in Table \ref{tab5b}, show a (small) performance improvement of the algorithms when Heuristic 2 is applied with a relative small increase of computational time. In fact, with Heuristic 1, the smallest average minimum gap is 3.59\%, found by Algorithm 13 (see Table  \ref{tab5}), while with Heuristic 2, the average minimum gap is 3.33\%, found by Algorithm 17 (see Table  \ref{tab5b}).

  Algorithms 13, 15 and 17 have been tested on the small and big instances of the third set, too. Tables \ref{tabsmall} and \ref{tabsmallb}  report the results of the algorithms on the 10 small instances starting from a solution provided by Heuristic 1 and Heuristic 2, respectively.
 In the tables,   for each instance,  ``av\_gap'' (``min\_gap') is the gap between the average (minimum) objective function value, ``av\_cost''   (``min\_cost''),      on the 5 best runs and the optimal solution ``opt", provided by CPLEX 12.5 running on the MILP formulation with a time limit of 1 hour.  ``av\_gap'' (``min\_gap') is computed as $(av\_cost-opt)/opt\times 100$ (as $(min\_cost-opt)/opt\times 100$).  
  In the first column of Tables \ref{tabsmall} and \ref{tabsmallb}, a * means that CPLEX was not able to certify the optimality of the solution within the time limit, while  a ** indicates that CPLEX was not able to find any feasible solution within the time limit.
  
Tables \ref{tabsmall} and \ref{tabsmallb} show Heuristic 2 gives overall slightly better results than Heuristic 1 even though the computational time is higher. The tables also show that Algorithm 13 works better than 15 and 17 with both heuristics. On the  instance 10, with 35 requests and 20 vehicles, our algorithms were able to find solutions  better than the solution found by CPLEX within the time limit. Observe that, the minimum gap is always smaller than 3.7\% and 2.3\% in Tables \ref{tabsmall} and \ref{tabsmallb}, respectively.
 
Finally, Tables \ref{tabbig} and \ref{tabbigb}  report the results of the algorithms on the 14 big instances of the third set, starting from a solution provided by Heuristic 1 and Heuristic 2, respectively. 
On these instances no feasible solution was found by CPLEX running on the MILP formulation with a time limit of 6 hours. 
These tables also show that on average Heuristic 2 works slightly better than Heuristic 1, in Algorithms 13 and 15, but perform slightly worse in Algorithm 17. Comparing the algorithms that employ Heuristic 1 (Table \ref{tabbig}), Algorithm 17 performs better than the others. On the other hand, in Table \ref{tabbig}, although  Algorithm 13 performs slightly worse in terms of ``av\_cost'' and time, it outperforms the others in terms of ``min\_cost''.

\begin{table}
\centering

\begin{tabular}{|ccccccc|}
\hline
Algo. \#        & 13         & 14   & 15             & 16  &17  &18 \\ \hline
Swap         & X         & X    & X              & X & X   & X \\ \hline
Rep Move     & X         & X    & X              & X  & X  & X\\ \hline
Move         & X         & X    & X              & X    & X& X\\ \hline
Chain        & X         &      & X              &      & X&\\ \hline
LS 3 & X         & X    &                &      & &\\ \hline
Focus LS      &           & X    & X              & X& &     \\ \hline
Focus LS with 8-step      &           &     &               & & X& X    \\ \hline
MaxSize      & 4         & 4    & 4              & 4  & 4  & 4\\ \hline
\end{tabular}
\caption{Algorithms of  the second test phase.}\label{tab:algo2}
\end{table}



\begin{sidewaystable*}
\scriptsize
\begin{center}

\begin{tabular}{|c|ccccc|ccccc|ccccc|}
\hline
  Algo.    & \multicolumn{5}{|c|}{ 13 }         & \multicolumn{5}{|c|}{14}          & \multicolumn{5}{|c|}{ 15 }     \\ \hline
 & av\_cost  &  min\_cost        & time       &av\_gap&min\_gap& av\_cost  &  min\_cost        & time       &av\_gap&min\_gap& av\_cost  &  min\_cost        & time       &av\_gap&min\_gap\\ \hline
a2-16&284.18&284.18&8.76&0.00&0.00&284.98&284.18&9.00&0.28&0.00&284.18&284.18&4.86&0.00&0.00\\ \hline
a2-20&359.64&359.64&24.07&0.21&0.21&359.34&358.88&29.22&0.13&0.00&359.13&358.88&16.19&0.07&0.00\\ \hline
a2-24&445.32&439.29&44.12&1.37&0.00&450.59&439.29&48.26&2.57&0.00&448.88&439.29&24.37&2.18&0.00\\ \hline
a3-18&296.41&292.56&26.74&1.37&0.05&297.70&297.38&30.40&1.81&1.70&297.50&297.38&17.26&1.74&1.70\\ \hline
a3-24&350.43&348.54&59.89&0.54&0.00&360.21&351.09&61.57&3.35&0.73&353.30&350.69&29.77&1.37&0.62\\ \hline
a3-30&503.73&487.09&8.17&3.64&0.22&508.77&494.24&9.97&4.68&1.69&503.12&490.65&4.70&3.51&0.95\\ \hline
a3-36&663.84&654.10&60.02&5.88&4.33&670.16&653.24&64.20&6.89&4.19&653.42&648.74&33.02&4.22&3.47\\ \hline
a4-16&285.54&285.40&14.33&0.05&0.00&285.40&285.40&14.88&0.00&0.00&285.40&285.40&7.98&0.00&0.00\\ \hline
a4-24&362.78&361.27&11.67&1.48&1.05&366.08&360.18&13.77&2.40&0.75&363.68&362.74&7.70&1.73&1.46\\ \hline
a4-32&477.31&477.17&66.32&1.22&1.19&478.20&477.24&75.24&1.41&1.21&477.61&476.24&47.26&1.29&1.00\\ \hline
a4-40&568.39&555.40&116.83&4.76&2.37&568.95&563.15&132.16&4.86&3.79&569.12&556.32&74.20&4.90&2.54\\ \hline
a4-48&663.86&658.35&204.88&4.12&3.26&665.38&660.30&244.14&4.36&3.56&666.70&663.99&120.77&4.57&4.14\\ \hline
a5-40&510.13&508.10&155.54&2.78&2.37&512.33&509.74&181.91&3.22&2.70&511.63&509.14&87.90&3.08&2.58\\ \hline
a5-50&703.34&695.75&295.82&5.09&3.95&701.03&689.44&339.76&4.74&3.01&700.27&693.36&172.46&4.63&3.59\\ \hline
a5-60&832.61&822.71&431.13&4.06&2.83&846.61&831.41&475.93&5.81&3.91&859.79&856.91&236.81&7.46&7.10\\ \hline
a6-48&628.93&612.08&146.45&7.31&4.44&628.59&625.54&210.96&7.25&6.73&630.33&605.81&104.19&7.55&3.37\\ \hline
a6-60&835.35&829.25&270.23&6.22&5.45&858.93&843.39&316.25&9.22&7.25&859.02&853.33&160.51&9.23&8.51\\ \hline
a6-72&975.16&964.48&622.27&7.13&5.96&982.64&971.48&673.56&7.95&6.73&972.32&950.10&351.48&6.82&4.38\\ \hline
a7-56&734.44&725.84&196.96&6.67&5.42&733.40&722.42&213.83&6.52&4.92&732.91&724.23&101.65&6.45&5.19\\ \hline
a7-70&952.35&950.34&393.29&7.45&7.22&950.82&944.88&376.47&7.28&6.60&941.57&930.97&231.69&6.23&5.04\\ \hline
a7-84&1180.57&1160.68&45.58&16.30&14.34&1165.91&1143.86&49.19&14.86&12.68&1170.50&1141.91&27.07&15.31&12.49\\ \hline
a8-64&784.81&763.91&204.45&10.05&7.12&777.50&767.84&241.63&9.03&7.67&780.61&768.30&126.45&9.47&7.74\\ \hline
a8-80&1007.20&984.55&440.86&8.82&6.37&995.07&987.61&431.47&7.51&6.70&1000.21&985.41&231.59&8.06&6.47\\ \hline
a8-96&1303.22&1293.59&661.48&8.93&8.13&1314.37&1289.01&779.91&9.87&7.75&1308.06&1284.10&357.96&9.34&7.33\\ \hline
Av   &654.56&646.43&187.91&4.81&3.59&656.79&647.97&209.32&5.25&3.93&655.39&646.59&107.41&4.97&3.74\\ \hline
\end{tabular}
\end{center}
\caption{ Performances of Algorithms 13-15 on the instances of the second set. }
\label{tab4}
\end{sidewaystable*}

\begin{sidewaystable*}
 \scriptsize
\begin{center}
\begin{tabular}{|c|ccccc|ccccc|ccccc|}
\hline
 Algo.     &  \multicolumn{5}{|c|}{16}             & \multicolumn{5}{|c|}{17 )}   &\multicolumn{5}{|c|}{18 }    \\ \hline
     & av\_cost   &  min\_cost        & time       &av\_gap&min\_gap& av\_cost   &  min\_cost        & time        &av\_gap&min\_gap& av\_cost   &  min\_cost        & time        &av\_gap&min\_gap\\ \hline
a2-16&284.18&284.18&4.52&0.00&0.00&284.18&284.18&5.62&0.00&0.00&285.78&284.18&14.96&0.56&0.00\\ \hline
a2-20&358.88&358.88&15.79&0.00&0.00&358.98&358.88&19.12&0.03&0.00&358.88&358.88&26.79&0.00&0.00\\ \hline
a2-24&451.04&439.29&23.81&2.67&0.00&439.29&439.29&28.83&0.00&0.00&439.29&439.29&43.23&0.00&0.00\\ \hline
a3-18&297.70&297.70&17.12&1.81&1.81&299.11&297.38&21.11&2.29&1.70&297.70&297.70&10.73&1.81&1.81\\ \hline
a3-24&359.07&351.09&27.90&3.02&0.73&358.61&348.54&37.14&2.89&0.00&357.47&348.93&16.98&2.56&0.11\\ \hline
a3-30&505.56&493.96&4.11&4.02&1.63&510.42&499.43&5.29&5.02&2.76&495.20&490.65&38.88&1.88&0.95\\ \hline
a3-36&673.12&658.53&30.53&7.36&5.04&656.14&648.52&34.17&4.65&3.44&644.73&634.23&52.85&2.83&1.16\\ \hline
a4-16&285.40&285.40&7.35&0.00&0.00&285.40&285.40&8.65&0.00&0.00&286.25&285.40&4.73&0.30&0.00\\ \hline
a4-24&364.86&362.74&7.29&2.06&1.46&361.76&357.51&9.33&1.19&0.00&362.43&360.18&13.51&1.38&0.75\\ \hline
a4-32&477.38&477.24&43.71&1.24&1.21&477.97&476.24&61.43&1.36&1.00&479.58&477.67&28.78&1.70&1.30\\ \hline
a4-40&569.85&564.23&68.85&5.03&3.99&558.29&556.51&85.36&2.90&2.57&563.17&555.50&39.06&3.80&2.38\\ \hline
a4-48&663.51&652.53&110.22&4.07&2.34&662.25&659.99&137.44&3.87&3.51&667.50&664.29&61.32&4.69&4.19\\ \hline
a5-40&510.82&508.93&92.78&2.91&2.53&511.07&507.29&115.94&2.96&2.20&517.70&512.42&29.73&4.30&3.24\\ \hline
a5-50&699.59&686.80&152.96&4.53&2.62&698.88&691.05&194.48&4.42&3.25&715.38&704.05&52.09&6.89&5.19\\ \hline
a5-60&852.04&842.18&224.59&6.49&5.26&844.27&832.50&292.58&5.52&4.05&876.24&862.89&81.09&9.52&7.85\\ \hline
a6-48&628.11&622.44&100.26&7.17&6.20&624.52&617.16&129.59&6.56&5.30&634.89&620.56&43.35&8.33&5.88\\ \hline
a6-60&851.82&842.67&156.64&8.32&7.16&851.94&837.80&200.11&8.33&6.54&870.20&862.54&55.25&10.66&9.68\\ \hline
a6-72&965.71&952.92&303.72&6.09&4.69&994.33&984.50&352.20&9.24&8.16&1010.78&1000.08&95.71&11.05&9.87\\ \hline
a7-56&731.23&719.14&104.76&6.20&4.45&729.53&717.08&119.11&5.96&4.15&737.11&729.27&38.76&7.06&5.92\\ \hline
a7-70&953.66&937.16&192.63&7.59&5.73&947.13&912.11&213.66&6.86&2.91&962.77&948.15&65.09&8.62&6.97\\ \hline
a7-84&1181.38&1168.73&24.74&16.38&15.13&1177.87&1164.48&29.22&16.03&14.71&1126.25&1111.81&117.07&10.95&9.53\\ \hline
a8-64&773.22&752.13&108.85&8.43&5.47&779.36&766.27&146.53&9.29&7.45&803.44&771.07&43.85&12.67&8.13\\ \hline
a8-80&995.84&985.29&192.75&7.59&6.45&996.99&983.64&244.03&7.72&6.27&1014.86&997.91&71.61&9.65&7.82\\ \hline
a8-96&1309.90&1286.87&341.47&9.49&7.57&1292.27&1288.36&465.11&8.02&7.69&1330.38&1318.17&111.49&11.20&10.18\\ \hline
Av   &655.99&647.13&98.22&5.10&3.81&654.19&646.42&123.17&4.80&3.65&659.92&651.49&48.20&5.52&4.29\\ \hline
\end{tabular}

\end{center}
\caption{ Performances of Algorithms 16-18 on the instances of the second set. }
\label{tab5}
\end{sidewaystable*}

\begin{sidewaystable*}
 \scriptsize
\begin{center}
\begin{tabular}{|c|ccccc|ccccc|ccccc|}
\hline
  Algo.    &  \multicolumn{5}{|c|}{13}             & \multicolumn{5}{|c|}{15 }   &\multicolumn{5}{|c|}{17 }    \\ \hline
     & av\_cost   &  min\_cost        & time       &av\_gap&min\_gap& av\_cost   &  min\_cost        & time        &av\_gap&min\_gap& av\_cost   &  min\_cost        & time        &av\_gap&min\_gap\\ \hline
a2-16&284.18&284.18&30.88&0.00&0.00&284.18&284.18&18.16&0.00&0.00&284.18&284.18&21.10&0.00&0.00\\ \hline
a2-20&359.64&359.64&42.96&0.21&0.21&359.49&358.88&28.26&0.17&0.00&359.19&358.88&33.10&0.09&0.00\\ \hline
a2-24&446.37&439.29&78.21&1.61&0.00&452.04&439.29&51.42&2.90&0.00&439.29&439.29&57.50&0.00&0.00\\ \hline
a3-18&298.10&298.01&13.62&1.95&1.91&296.76&292.41&10.95&1.49&0.00&296.89&292.41&11.82&1.53&0.00\\ \hline
a3-24&350.86&350.69&37.84&0.67&0.62&353.18&350.69&24.37&1.33&0.62&349.14&348.54&27.08&0.17&0.00\\ \hline
a3-30&489.86&487.68&75.18&0.79&0.34&488.81&487.09&52.87&0.57&0.22&489.22&487.89&56.37&0.66&0.38\\ \hline
a3-36&647.17&647.17&122.16&3.22&3.22&-&-&-&-&-&649.50&641.63&97.41&3.59&2.34\\ \hline
a4-16&285.69&285.40&8.27&0.10&0.00&285.40&285.40&5.71&0.00&0.00&285.40&285.40&6.26&0.00&0.00\\ \hline
a4-24&363.07&361.70&25.04&1.56&1.17&362.51&361.72&18.39&1.40&1.18&362.35&361.72&21.14&1.35&1.18\\ \hline
a4-32&480.03&476.71&52.17&1.80&1.10&484.75&482.45&42.79&2.80&2.31&481.82&477.01&47.62&2.18&1.16\\ \hline
a4-40&565.45&555.50&100.92&4.22&2.38&572.59&567.10&80.07&5.54&4.52&563.01&550.71&84.64&3.77&1.50\\ \hline
a4-48&657.81&649.52&192.83&3.17&1.87&664.26&656.14&148.43&4.18&2.91&659.15&654.19&153.06&3.38&2.61\\ \hline
a5-40&516.19&512.23&71.84&3.99&3.20&518.36&511.51&55.80&4.43&3.05&516.53&510.76&60.52&4.06&2.90\\ \hline
a5-50&704.31&694.32&177.18&5.23&3.74&698.14&687.43&124.53&4.31&2.71&704.35&696.11&135.27&5.24&4.01\\ \hline
a5-60&854.52&847.57&289.22&6.80&5.93&860.72&857.17&226.44&7.58&7.13&858.33&845.59&239.65&7.28&5.69\\ \hline
a6-48&623.17&612.68&103.88&6.33&4.54&629.43&625.23&83.29&7.40&6.68&632.94&618.84&95.16&8.00&5.59\\ \hline
a6-60&845.15&814.04&201.95&7.47&3.51&851.21&837.25&161.99&8.24&6.47&848.80&820.90&174.55&7.93&4.39\\ \hline
a6-72&969.86&954.76&397.47&6.55&4.89&964.20&944.42&326.36&5.93&3.75&958.21&936.54&337.31&5.27&2.89\\ \hline
a7-56&745.45&736.65&136.79&8.27&6.99&739.86&732.72&116.03&7.46&6.42&741.96&726.18&117.29&7.76&5.47\\ \hline
a7-70&967.13&960.55&275.81&9.12&8.37&970.33&962.32&235.28&9.48&8.57&974.71&960.81&246.87&9.97&8.40\\ \hline
a7-84&1110.65&1076.09&562.09&9.41&6.01&1106.21&1091.93&498.67&8.97&7.57&1098.81&1087.15&517.91&8.25&7.10\\ \hline
a8-64&798.65&790.30&177.08&11.99&10.82&788.61&778.99&149.35&10.59&9.24&780.33&763.54&159.63&9.43&7.07\\ \hline
a8-80&998.78&994.88&358.96&7.91&7.49&999.19&990.27&311.70&7.95&6.99&1003.53&993.35&314.21&8.42&7.32\\ \hline
a8-96&1324.05&1310.70&723.40&10.67&9.56&1326.35&1298.00&661.01&10.87&8.50&1322.03&1316.49&665.05&10.51&10.04\\ \hline
Av. &653.59&645.84&177.32&4.71&3.66&654.63&647.07&149.21&4.94&3.86&652.49&644.09&153.36&4.54&3.33\\ \hline
\end{tabular}
\end{center}
\caption{ Performances of Algorithms 13, 15 and 17 on the instances of the second set, starting from a solution provided by Heuristic 2. }
\label{tab5b}
\end{sidewaystable*}

\begin{sidewaystable*}
 \scriptsize
\begin{center}
\begin{tabular}{|c|cc|ccccc|ccccc|ccccc|}
\hline
  Algo.  &&  &  \multicolumn{5}{|c|}{13}             & \multicolumn{5}{|c|}{15 }   &\multicolumn{5}{|c|}{17 }    \\ \hline
     Inst.&$n$&$m$& av\_cost   &  min\_cost        & time       &av\_gap&min\_gap& av\_cost   &  min\_cost        & time        &av\_gap&min\_gap& av\_cost   &  min\_cost        & time        &av\_gap&min\_gap\\ \hline
1&10&4&252.00&252.00&1.41&0.00&0.00&252.00&252.00&1.07&0.00&0.00&252.00&252.00&1.30&0.00&0.00\\ \hline
2&20&14&859.20&854.00&2.09&2.41&1.79&868.60&856.00&1.66&3.53&2.03&868.60&856.00&1.52&3.53&2.03\\ \hline
3&20&15&539.80&538.00&1.89&1.47&1.13&541.20&538.00&1.57&1.73&1.13&541.20&538.00&1.55&1.73&1.13\\ \hline
4*&22&16&878.40&876.00&2.47&2.26&1.98&895.20&890.00&2.06&4.21&3.61&895.20&890.00&2.07&4.21&3.61\\ \hline
5&25&18&733.80&729.00&2.30&1.49&0.83&733.80&732.00&2.01&1.49&1.24&736.20&732.00&2.12&1.83&1.24\\ \hline
6&25&18&677.40&661.00&2.36&3.90&1.38&674.60&667.00&1.86&3.47&2.30&674.60&667.00&1.98&3.47&2.30\\ \hline
7*&30&20&844.20&830.00&2.93&1.83&0.12&863.20&847.00&2.53&4.13&2.17&856.00&847.00&2.86&3.26&2.17\\ \hline
8*&30&20&987.40&979.00&2.55&3.18&2.30&1003.80&968.00&2.30&4.89&1.15&983.20&968.00&2.48&2.74&1.15\\ \hline
9**&35&20&912.80&904.00&3.39&-&-&932.80&920.00&3.00&-&-&925.60&914.00&3.39&-&-\\ \hline
10*&35&20&1061.20&1036.00&2.97&-0.17&-2.54&1059.00&1047.00&2.50&-0.38&-1.51&1062.40&1047.00&2.98&-0.06&-1.51\\ \hline
 Av.& & &774.62&765.90&2.43&1.82&0.78&782.42&771.70&2.06&2.56&1.35&779.50&771.10&2.23&2.30&1.35\\ \hline
\end{tabular}
\end{center}
\caption{Performances of Algorithms 13, 15 and 17 on the small instances of the third set, starting from a solution provided by Heuristic 1.}
\label{tabsmall}
\end{sidewaystable*}

\begin{sidewaystable*}
 \scriptsize
\begin{center}
\begin{tabular}{|c|cc|ccccc|ccccc|ccccc|}
\hline
 Algo.   &&  &  \multicolumn{5}{|c|}{13}             & \multicolumn{5}{|c|}{15 }   &\multicolumn{5}{|c|}{17 }    \\ \hline
     Inst.&$n$&$m$& av\_cost   &  min\_cost        & time       &av\_gap&min\_gap& av\_cost   &  min\_cost        & time        &av\_gap&min\_gap& av\_cost   &  min\_cost        & time        &av\_gap&min\_gap\\ \hline
1&10&4&252.00&252.00&1.67&0.00&0.00&252.00&252.00&1.31&0.00&0.00&252.00&252.00&1.33&0.00&0.00\\ \hline
2&20&14&849.60&842.00&2.92&1.26&0.36&852.40&842.00&2.46&1.60&0.36&852.40&842.00&2.53&1.60&0.36\\ \hline
3&20&15&541.80&538.00&2.54&1.84&1.13&543.40&538.00&2.13&2.14&1.13&543.40&538.00&2.16&2.14&1.13\\ \hline
4*&22&16&877.80&876.00&4.01&2.19&1.98&876.00&864.00&3.06&1.98&0.58&876.00&864.00&3.35&1.98&0.58\\ \hline
5&25&18&738.80&736.00&3.64&2.19&1.80&736.40&732.00&3.09&1.85&1.24&734.40&729.00&3.13&1.58&0.83\\ \hline
6&25&18&674.80&667.00&4.27&3.50&2.30&671.00&657.00&3.55&2.91&0.77&671.00&657.00&3.60&2.91&0.77\\ \hline
7*&30&20&845.60&834.00&5.66&2.00&0.60&852.60&848.00&4.92&2.85&2.29&852.60&848.00&5.02&2.85&2.29\\ \hline
8*&30&20&977.60&972.00&5.25&2.15&1.57&982.20&969.00&4.52&2.63&1.25&982.20&969.00&4.53&2.63&1.25\\ \hline
9**&35&20&913.80&908.00&8.77&-&-&927.20&912.00&7.96&-&-&931.20&912.00&7.87&-&-\\ \hline
10*&35&20&1064.80&1046.00&7.17&0.17&-1.60&1058.60&1046.00&6.36&-0.41&-1.60&1058.60&1046.00&6.30&-0.41&-1.60\\ \hline
 Av.& & &773.66&767.10&4.59&1.70&0.90&775.18&766.00&3.94&1.73&0.67&775.38&765.70&3.98&1.70&0.62\\ \hline
\end{tabular}
\end{center}
\caption{Performances of Algorithms 13, 15 and 17 on the small instances of the third set, starting from a solution provided by Heuristic 2.}
\label{tabsmallb}
\end{sidewaystable*}

\begin{sidewaystable*}
 \scriptsize
\begin{center}
\begin{tabular}{|c|cc|ccc|ccc|ccc|}
\hline
 Algo.   &&  &  \multicolumn{3}{|c|}{13}             & \multicolumn{3}{|c|}{15 }   &\multicolumn{3}{|c|}{17 }    \\ \hline
     Inst.&$n$&$m$& av\_cost   &  min\_cost        & time      &av\_cost   &  min\_cost        & time      &av\_cost   &  min\_cost        & time      \\ \hline
1&80&60&1932.20&1908.00&35.41&1931.20&1926.00&33.75&1938.40&1912.00&33.96 \\ \hline
2&80&60&1910.60&1896.00&34.19&1923.60&1897.00&33.52&1917.40&1897.00&34.34 \\ \hline
3&80&60&2022.20&2005.00&33.31&2018.20&2001.00&33.53&2024.00&1994.00&33.43 \\ \hline
4&80&60&2195.00&2137.00&33.71&2218.00&2134.00&33.87&2204.20&2149.00&33.86 \\ \hline
5&100&75&2630.00&2619.00&49.76&2618.60&2574.00&48.98&2635.20&2619.00&48.82 \\ \hline
6&100&75&3106.00&3094.00&50.37&3123.20&3070.00&49.30&3114.00&3070.00&49.35 \\ \hline
7&100&75&2718.40&2677.00&49.33&2719.20&2658.00&48.62&2720.80&2658.00&48.33 \\ \hline
8&100&75&2705.80&2671.00&49.95&2721.80&2698.00&48.82&2720.60&2687.00&49.06 \\ \hline
9&100&75&2857.40&2778.00&49.18&2920.20&2790.00&48.61&2818.40&2790.00&49.17 \\ \hline
10&100&80&3233.20&3206.00&53.84&3260.40&3196.00&50.76&3221.60&3190.00&50.53 \\ \hline
11&100&80&2912.20&2908.00&51.96&2989.80&2868.00&48.63&2913.20&2900.00&49.72 \\ \hline
12&100&80&2879.20&2858.00&50.99&2859.80&2847.00&49.90&2844.20&2827.00&51.18 \\ \hline
13&100&80&2864.00&2830.00&52.75&2810.40&2794.00&49.73&2818.00&2794.00&50.89 \\ \hline
14&100&80&2696.00&2687.00&51.97&2679.00&2664.00&48.60&2679.00&2664.00&49.88 \\ \hline
Av.& &&2618.73&2591.00&46.19&2628.10&2579.79&44.76&2612.07&2582.21&45.18\\ \hline
\end{tabular}
\end{center}
\caption{Performances of Algorithms 13, 15 and 17 on the big instances of the third set, starting from a solution provided by Heuristic 1.}
\label{tabbig}
\end{sidewaystable*}

\begin{sidewaystable*}
 \scriptsize
\begin{center}
\begin{tabular}{|c|cc|ccc|ccc|ccc|}
\hline
 Algo.   &&  &  \multicolumn{3}{|c|}{13}             & \multicolumn{3}{|c|}{15 }   &\multicolumn{3}{|c|}{17 }    \\ \hline
     Inst.&$n$&$m$& av\_cost   &  min\_cost        & time      &av\_cost   &  min\_cost        & time      &av\_cost   &  min\_cost        & time      \\ \hline1&80&60&1918.20&1905.00&73.06&1916.80&1895.00&69.55&1915.80&1904.00&70.14\\ \hline
2&80&60&1911.40&1871.00&72.32&1898.20&1868.00&69.69&1909.80&1868.00&70.26\\ \hline
3&80&60&2025.20&2004.00&68.23&2019.40&1994.00&66.73&2004.00&1970.00&66.94\\ \hline
4&80&60&2208.40&2161.00&68.97&2197.40&2175.00&66.96&2194.80&2175.00&67.56\\ \hline
5&100&75&2660.20&2650.00&112.73&2628.40&2593.00&110.98&2662.40&2627.00&109.85\\ \hline
6&100&75&3138.20&3126.00&111.95&3120.40&3083.00&113.15&3135.80&3121.00&109.26\\ \hline
7&100&75&2704.40&2676.00&109.28&2711.80&2688.00&108.40&2710.60&2695.00&107.11\\ \hline
8&100&75&2700.80&2668.00&112.13&2730.40&2711.00&112.31&2686.20&2669.00&109.56\\ \hline
9&100&75&2845.00&2818.00&107.85&2813.20&2788.00&106.33&2823.20&2788.00&106.00\\ \hline
10&100&80&3252.80&3206.00&110.12&3254.80&3229.00&107.87&3263.60&3246.00&107.81\\ \hline
11&100&80&2913.40&2867.00&107.80&2903.60&2880.00&108.58&2901.20&2880.00&109.83\\ \hline
12&100&80&2849.20&2792.00&108.09&2866.80&2832.00&108.98&2903.60&2867.00&109.77\\ \hline
13&100&80&2839.20&2791.00&111.41&2830.60&2810.00&112.43&2846.40&2829.00&110.84\\ \hline
14&100&80&2657.00&2601.00&110.09&2653.00&2617.00&111.28&2661.20&2651.00&109.76\\ \hline
Av.& & &2615.96&2581.14&98.86&2610.34&2583.07&98.09&2615.61&2592.14&97.48\\ \hline
\end{tabular}
\end{center}
\caption{Performances of Algorithms 13, 15 and 17 on the big instances of the third set, starting from a solution provided by Heuristic 2.}
\label{tabbigb}
\end{sidewaystable*}

\subsection{Results of the second test phase on single-depot instances}\label{sec:resparragh}
In Tables \ref{tab4c} and  \ref{tab4b} results of Algorithms 13 and 15 running on the single-depot instances of Set 1 are reported. In the tables, the gap values are computed with the best known solution (given in \cite{Braekers}). A ``-'' in the columns reporting the gaps  indicates that no solution was available for the instance. 

Observe that, Algorithm 15 often attains better results in terms of solution quality with computational times higher of about 20\%. Furthermore, note that, in general, the hardest instances are those with the highest $n/m$ ratio, i.e., a3-36, a4-40 and a4-48. When comparing to the best known results, Algorithm 15 finds better results on the instances with homogeneous vehicles and users followed by the instances with heterogeneous vehicles and users. Algorithm 13 instead, finds better solutions in homogeneous vehicles and heterogeneous users followed by the instances with heterogeneous vehicles and users.

\begin{sidewaystable*}
 \scriptsize

\begin{center}

\begin{tabular}{|l|l|l|l|l|l|l|l|l|l|l|l|l|l|l|l|}
\hline
      & \multicolumn{5}{c|}{U}                         & \multicolumn{5}{c|}{I}                         & \multicolumn{5}{c|}{E}                         \\ \hline
      & cost    &  min\_cost     & time   & av\_gap & min\_gap & cost    &  min\_cost     & time   & av\_gap & min\_gap & cost    &  min\_cost     & time   & av\_gap & min\_gap \\ \hline
a2-16 & 297.14  & 294.25  & 21.93  & 0.98   & 0.00       & 333.53  & 331.16  & 23.10  & 0.72   & 0.00       & 297.06  & 294.25  & 21.85  & 0.96   & 0.00       \\ \hline
a2-20 & 349.68  & 344.83  & 42.03  & 1.41   & 0.00       & 351.79  & 349.29  & 39.60  & 1.37   & 0.65       & 359.79  & 357.99  & 47.81  & 1.14   & 0.63       \\ \hline
a2-24 & 442.01  & 431.22  & 67.88  & 2.53   & 0.02       & 454.71  & 450.35  & 67.26  & 0.99   & 0.02       & 440.40  & 431.22  & 67.17  & 2.15   & 0.02       \\ \hline
a3-18 & 304.88  & 300.72  & 13.03  & 1.46   & 0.08       & 305.99  & 300.87  & 13.59  & 1.78   & 0.08       & 304.88  & 302.41  & 12.83  & 0.90   & 0.08       \\ \hline
a3-24 & 350.68  & 345.23  & 28.78  & 1.70   & 0.12       & 348.39  & 345.40  & 27.00  & 1.01   & 0.14       & 349.33  & 345.33  & 26.20  & 1.31   & 0.14       \\ \hline
a3-30 & 499.99  & 495.26  & 58.38  & 1.04   & 0.08       & 507.07  & 500.99  & 58.56  & 1.30   & 0.08       & 503.97  & 495.97  & 57.38  & 1.84   & 0.23       \\ \hline
a3-36 & 624.65  & 614.54  & 81.21  & 7.11   & 5.38       & 621.38  & 597.43  & 80.82  & 6.55   & 2.44       & 657.74  & 647.31  & 84.62  & 6.33   & 4.65       \\ \hline
a4-16 & 285.08  & 282.68  & 8.49   & 0.85   & 0.00       & 294.16  & 291.55  & 7.32   & 2.86   & 1.94       & 302.65  & 301.81  & 9.20   & 1.20   & 0.92       \\ \hline
a4-24 & 381.97  & 375.32  & 21.33  & 1.85   & 0.08       & 388.13  & 383.89  & 20.67  & 1.12   & 0.01       & 382.16  & 376.86  & 20.53  & 1.90   & 0.49       \\ \hline
a4-32 & 491.69  & 490.22  & 30.28  & 1.27   & 0.97       & 506.16  & 503.18  & 31.79  & 0.72   & 0.00       & 493.49  & 491.34  & 30.83  & 1.35   & 0.91       \\ \hline
a4-40 & 596.47  & 583.07  & 63.83  & 6.95   & 4.55       & 613.57  & 601.96  & 59.73  & 4.77   & 0.03       & 595.61  & 579.47  & 63.13  & 6.80   & 3.90       \\ \hline
a4-48 & 692.70  & 685.49  & 120.38 & 2.83   & 1.76       & 693.70  & 678.38  & 114.15 & 2.71   & 0.00       & 700.77  & 688.70  & 117.33 & 2.99   & 1.22       \\ \hline
a5-40 & 527.12  & 510.03  & 45.01  &      -  & -           & 525.51  & 509.38  & 46.59  &   -     &      -      & 530.80  & 514.14  & 49.10  &    -    &     -       \\ \hline
a5-50 & 714.83  & 698.83  & 109.56 &        -  & -            & 726.72  & 720.40  & 111.23 &   -     &      -      & 730.29  & 707.01  & 96.43  &    -    &     -       \\ \hline
a5-60 & 880.64  & 840.77  & 147.19 &        -  & -            & 879.85  & 855.32  & 136.54 &   -     &      -      & 887.61  & 861.33  & 139.76 &    -    &     -       \\ \hline
a6-48 & 627.67  & 618.08  & 71.77  &        -  & -            & 643.43  & 630.71  & 63.37  &   -     &      -      & 635.71  & 619.56  & 55.61  &    -    &     -       \\ \hline
a6-60 & 892.08  & 848.09  & 88.50  &        -  & -            & 910.61  & 894.73  & 82.63  &   -     &      -      & 901.72  & 863.59  & 84.39  &    -    &     -       \\ \hline
a6-72 & 980.24  & 953.26  & 175.44 &        -  & -            & 1025.28 & 996.93  & 165.57 &   -     &      -      & 1010.25 & 988.58  & 184.87 &    -    &     -       \\ \hline
a7-56 & 777.52  & 757.48  & 64.19  &   -     &      -      & 784.80  & 766.06  & 60.59  &   -     &      -      & 781.98  & 754.98  & 62.53  &    -    &     -       \\ \hline
a7-70 & 983.83  & 960.90  & 98.27  &   -     &      -      & 1010.72 & 992.68  & 90.28  &   -     &      -      & 1028.56 & 983.84  & 94.75  &    -    &     -       \\ \hline
a7-84 & 1117.60 & 1082.47 & 194.86 &   -     &      -      & 1178.47 & 1146.17 & 171.75 &   -     &      -      & 1130.35 & 1105.59 & 191.77 &    -    &     -       \\ \hline
a8-64 & 804.61  & 795.56  & 85.93  &   -     &      -      & 828.24  & 816.93  & 79.00  &   -     &      -      & 816.17  & 798.06  & 70.40  &    -    &     -       \\ \hline
a8-80 & 1017.43 & 991.59  & 149.14 &   -     &      -      & 1066.62 & 1049.00 & 138.53 &   -     &      -      & 1046.38 & 1033.41 & 130.17 &    -    &     -       \\ \hline
a8-96 & 1332.76 & 1307.82 & 182.43 &   -     &      -      & 1377.58 & 1351.00 & 194.88 &   -     &      -      & 1378.21 & 1340.98 & 198.28 &    -    &     -       \\ \hline
Av.    & 665.55  & 650.32  & 82.08  & 2.50   & 1.09       & 682.35  & 669.32  & 78.52  & 2.16   & 0.45       & 677.74  & 661.82  & 79.87  & 2.41   & 1.10       \\ \hline
\end{tabular}
\end{center}
\caption{ Performance of Algorithms 13 on the single depot instances of Set 1, starting  solution provided by Heuristic 1}
\label{tab4c}
\end{sidewaystable*}

\begin{sidewaystable*}
 \scriptsize

\begin{center}

\begin{tabular}{|l|l|l|l|l|l|l|l|l|l|l|l|l|l|l|l|}
\hline
      & \multicolumn{5}{c|}{U}                         & \multicolumn{5}{c|}{I}                         & \multicolumn{5}{c|}{E}                         \\ \hline
      & cost    &  min\_cost     & time   & av\_gap & min\_gap & cost    &  min\_cost     & time   & av\_gap & min\_gap & cost    &  min\_cost     & time   & av\_gap & min\_gap \\ \hline
a2-16&296.44&294.25&26.59&0.74&0.00&334.35&333.95&25.78&0.96&0.84&297.66&295.19& 25.7731  .&1.16&0.32\\ \hline
a2-20&345.99&344.88&52.47&0.34&0.01&348.87&347.03&48.21&0.53&0.00&357.58&355.78&52.53&0.52&0.01\\ \hline
a2-24&432.48&431.22&80.39&0.32&0.02&455.22&450.35&84.66&1.10&0.02&435.24&431.22&78.03&0.96&0.02\\ \hline
a3-18&306.50&300.72&16.10&2.00&0.08&308.35&301.19&15.98&2.57&0.19&306.52&302.41&16.12&1.44&0.08\\ \hline
a3-24&349.15&345.23&31.85&1.25&0.12&349.12&347.26&31.52&1.22&0.68&350.88&345.23&31.01&1.75&0.12\\ \hline
a3-30&505.09&500.07&62.92&2.07&1.06&512.77&504.51&66.12&2.44&0.79&510.83&495.97&69.03&3.23&0.23\\ \hline
a3-36&610.30&604.64&102.61&4.65&3.68&619.81&598.88&97.53&6.28&2.69&641.74&626.60&98.42&3.74&1.30\\ \hline
a4-16&286.60&283.10&10.89&1.39&0.15&294.39&291.55&8.69&2.94&1.94&302.65&301.81&9.06&1.20&0.92\\ \hline
a4-24&382.18&375.32&25.61&1.91&0.08&389.93&384.14&22.62&1.59&0.08&382.18&377.11&23.30&1.91&0.56\\ \hline
a4-32&493.17&490.56&38.67&1.58&1.04&506.27&503.18&38.10&0.75&0.13&496.00&491.67&40.30&1.86&0.97\\ \hline
a4-40&584.96&562.21&77.58&4.89&0.81&608.74&595.06&72.09&3.94&1.61&588.99&569.03&73.31&5.61&2.03\\ \hline
a4-48&695.73&690.05&138.64&4.02&3.17&695.57&682.84&138.23&2.99&1.11&705.97&693.06&139.21&3.75&1.86\\ \hline
a5-40&526.45&511.33&60.69&-&-&526.33&512.96&56.42&-&-&528.37&514.37&60.19&-&-\\ \hline
a5-50&726.87&714.05&129.75&-&-&732.91&718.80&119.55&-&-&725.85&708.16&125.08&-&-\\ \hline
a5-60&866.13&830.58&171.79&-&-&878.06&860.90&154.96&-&-&885.35&870.11&170.51&-&-\\ \hline
a6-48&629.05&618.52&76.48&-&-&643.82&630.55&78.69&-&-&633.61&618.31&68.28&-&-\\ \hline
a6-60&905.96&850.46&113.15&-&-&909.57&880.47&102.36&-&-&887.54&859.13&101.78&-&-\\ \hline
a6-72&980.77&963.86&194.90&-&-&1022.72&1006.55&188.37&-&-&1000.50&972.96&184.88&-&-\\ \hline
a7-56&776.24&757.65&72.77&-&-&793.91&786.22&66.43&-&-&777.21&755.82&80.72&-&-\\ \hline
a7-70&982.44&971.46&129.44&-&-&1016.25&989.49&112.17&-&-&1000.60&975.73&115.33&-&-\\ \hline
a7-84&1138.53&1101.51&230.05&-&-&1173.09&1148.49&208.16&-&-&1128.27&1091.93&232.08&-&-\\ \hline
a8-64&802.65&786.46&87.68&-&-&826.80&793.79&78.45&-&-&809.11&798.18&83.85&-&-\\ \hline
a8-80&1021.24&998.78&192.41&-&-&1068.53&1055.58&167.53&-&-&1041.92&1016.19&170.13&-&-\\ \hline
a8-96&1324.08&1305.76&239.61&-&-&1376.92&1328.29&229.18&-&-&1389.33&1344.84&243.29&-&-\\ \hline
Av.  &665.37&651.36&98.46&2.10&0.85&683.01&668.83&92.16&2.28&0.84&674.33&658.78&95.51&2.26&0.70\\ \hline
\end{tabular}
\end{center}
\caption{ Performance of Algorithms 15 on the single depot instances of Set 1, starting  solution provided by Heuristic 1. }
\label{tab4b}
\end{sidewaystable*}


\section{Conclusion}\label{sec:conc}

In this work, a study of Variable Neighborhood Search (VNS) algorithms for multi-depot dial-a-ride problems has been addressed. Different versions of Variable Neighborhood Search algorithm have been proposed able to tackle all the characteristics of the problem. The different versions of the VNS algorithms have been tested on literature instances and on 24 random instances taken from a real-life healthcare problem.

Future research includes the study of new neighborhood structures able to tackle with the heterogeneity and multi-depot issues. The development of new local search approaches to use in combination with the VNS technique.




\begin{thebibliography}{50}

\bibitem{Agnetis}
A. Agnetis,  G. De Pascale, P. Detti, J. Raffaelli, P. Chelli, R. Colombai, G. Marconcini, E. Porfido, and A. Coppi,
{Applicazione di tecniche di operations management per minimizzare il costo di trasporto di pazienti}, 
MECOSAN, Italian Quart. of Health Care Management Economics and Politics, 84, 2012.





\bibitem{Bard}
J. F. Bard, A. I. Jarrah, Integrating commercial and residential pickup and delivery networks: A case study, Omega, 41 (4),  706--720, 2013.

\bibitem{Beaudry}
A. Beaudry, G. Laporte, T. Melo, S. Nickel, Dynamic transportation of patients in hospitals, OR Spectrum, 32 (1), 77--107, 2010.

\bibitem{Berbeglia}
G. Berbeglia, J.-F. Cordeau, I. Gribkovskaia, G. Laporte, Static pickup and delivery problems: a classification scheme and survey. TOP,  15, 1--31, 2007.







\bibitem{Bettinelli}
A. Bettinelli, A. Ceselli, and G. Righini, 
\emph{A branch-and-cut-and-price algorithm for the multi-depot heterogeneous vehicle routing problem with time windows},
Transportation Research Part C, 19, 723--740, 2011.


\bibitem{Braekers}
K. Braekers,  A. Caris, G.K. Janssens,  Exact and metaheuristic approach for a general heterogeneous dial-a-ride problem with multiple depots. Transportation Research Part B, 67, 6--186, 2014.





\bibitem{Carnes}
T.A. Carnes, S.G. Henderson, D.B. Shmoys, M. Ahghari, R.D. MacDonald, Mathematical programming guides air-ambulance routing at ornge.
Interfaces, 43 (3), 232--239, 2013.

\bibitem{Cli}
J. C. N. Climaco, J. M. F. Craveirinha, and M. M. B. Pascoal. An automated reference point-like approach for multicriteria shortest path problems. Journal of Systems Science and Systems Engineering, 15, 314--329, 2006.

\bibitem{coppi}
A. Coppi, P. Detti, J. Raffaelli, A planning and routing model for patient transportation in healthcare, Electronic Notes in Discrete Mathematics, 41,  125--132, 2013.

\bibitem{Cor2006} J.-F. Cordeau,  A Branch-and-Cut Algorithm for the Dial-a-Ride Problem. Operations Research, 54(3), 573--586, 2006. 

\bibitem{CorLap3}
J.-F. Cordeau and  G. Laporte, A tabu search heuristic for the static multi-vehicle dial-a-ride problem. Transportation Research Part B, 37, 579--594,  2003.

\bibitem{CorLap}
J.-F. Cordeau and G. Laporte.  The dial-a-ride problem: models and algorithms. Annals of  Operations Research, 153:29--46, 2007.


\bibitem{mic2015}
P. Detti, F. Papalini, G. Zabalo Manrique de Lara, A multi-depot dial-a-ride problem for patients transportation in healthcare, Proceedings of the 11th Metaheuristic International Conference (MIC), Agadir,  7--10 June, 2015.

\bibitem{DPZomega}
P. Detti, F. Papalini, G. Zabalo Manrique de Lara, A multi-depot dial-a-ride problem with heterogeneous vehicles and compatibility constraints in healthcare, submitted.

\bibitem{Erdogan}
G. Erdogan, E. Erkut, A. Ingolfsson, G. Laporte, Scheduling ambulance crews for maximum coverage, J. Oper. Res. Soc. 61 (4), 543--550, 2010.



\bibitem{Hanne}
T. Hanne, T. Melo, S. Nickel, Bringing robustness to patient flow management through optimized patient transports in hospitals, Interfaces 39 (3), 241--255, 2009.


\bibitem{Liu}
R. Liu, X. Xie, T. Garaix, Hybridization of tabu search with feasible and infeasible local searches for periodic home health care logistics, Omega, 47,  17--32,  2014.

\bibitem{Ma}
H. Ma, B. Cheang, A. Lim, L. Zhang, Y. Zhu, An investigation into the vehicle routing problem with time windows and link capacity constraints, Omega, 40 (3), 336--347, 2012. 


\bibitem{Melachrinoudisa}
E. Melachrinoudis, A. B. Ilhana, and H. Min, 
{A dial-a-ride problem for client transportation in a health-care organization},  
Computers \& Operations Research, 34, 742--759, 2007.

\bibitem{Muelas}
S. Muelas, A. LaTorre, J.-M. Pena, A variable neighborhood search algorithm for the optimization of a dial-a-ride problem in a large city, Expert Systems with Applications,  40,  14, 5516--5531, 2013.


\bibitem{Paquette}
J. Paquette, J.-F. Cordeau, G. Laporte,  M.M.B. Pascoal, Combining multicriteria analysis and tabu search for dial-a-rial problems.Transportation Research B, 46, 100--119, 2012.

\bibitem{Parragh2011}
S.N. Parragh, Introducing heterogeneous users and vehicles into models and algorithms for the dial-a-ride problem. Transp. Res. Part C: Emerg. Technol. 19 (5), 912--930, 2011.

\bibitem{Parragh}
S. N. Parragh, K. F. Doerner, and R. F. Hartl, Variable neighborhood search for the dial-a-ride problem, Computers \& Operations Research, 37, 6, 1129--1138, 2010. 

\bibitem{Parragh1}
S. N. Parragh, J. F. Cordeau, K. F. Doerner, and R. F. Hartl, Models and algorithms for the heterogeneous dial-a-ride problem with driver-related constraints. OR Spectrum 34, 3 , 593--633, 2012.
\bibitem{Parragh2013}
S. N. Parragh and V. Schmid, Hybrid column generation and large neighborhood search for the dial-a-ride problem, Computers \& Operations Research, 40, 1, 490--497, 2013. 


\bibitem{Rekiek} 
B. Rekiek,  A. Delchambre, H.A. Saleh, Handicapped person transportation: an application of the grouping genetic algorithm, Engineering Applications of Artificial Intelligence, 19:511--520, 2006.

\bibitem{Ropke2007}
S. Ropke, J.F. Cordeau,  G. Laporte, G.,  Models and branch-and-cut algorithms for pickup and delivery problems with time windows. Networks 49 (4),
258--272, 2007.

\bibitem{Ropke}
S. Ropke  and J.F. Cordeau, 
{Branch and Cut and Price for the Pickup and Delivery Problem with Time Windows}, 
Transportation Science, 43, 267--286, 2009.

\bibitem{Savelsbergh} 
M. Savelsbergh and M. Sol, 
{DRIVE: Dynamic Routing of Independent Vehicles}, 
Operations Research, 46, 474--490, 1998. 


\end{thebibliography}


\end{document}